\documentclass{article}
\usepackage{spconf,amsmath,graphicx}
\usepackage{xcolor}
\usepackage{multirow}

\title{AutoKWS: Keyword Spotting with Differentiable  Architecture Search}
\name{Bo Zhang, Wenfeng Li, Qingyuan Li, Weiji Zhuang, Xiangxiang Chu, Yujun Wang} 

\address{Xiaomi AI Lab}
\begin{document}
%
\maketitle
\begin{abstract}
Smart audio devices are gated by an always-on lightweight keyword spotting program to reduce power consumption. It is however challenging to design models that have both high accuracy and low latency for accurate and fast responsiveness. Many efforts have been made to develop end-to-end neural networks, in which depthwise separable convolutions, temporal convolutions, and LSTMs are adopted as building units. Nonetheless, these networks designed with human expertise may not achieve an optimal trade-off in an expansive search space. In this paper, we propose to leverage recent advances in differentiable neural architecture search to discover more efficient networks. Our searched model attains 97.2\% top-1 accuracy on Google Speech Command Dataset v1 with only nearly 100K parameters.
\end{abstract}
\begin{keywords}
Keyword spotting, neural architecture search
\end{keywords}

\section{Introduction}
With the fast evolution of deep learning in audio systems, keyword spotting (KWS) plays an increasingly important role in smart device terminals like mobile phones and smart speakers. In real applications,  a keyword is predefined for the keyword detection system, such as  ``OK Google", ``Hey Siri", ``Xiao Ai Tong Xue". When the keyword is detected, the following audio stream can then be uploaded to speech recognition systems. The goal of KWS is to quickly and accurately detect keywords in real-time. Thus it is necessary to evaluate the model performance in three aspects: accuracy, the number of parameters, and computational cost (typically multiply-adds). In previous work, the algorithms for KWS are generally divided into four types: (a) feature template matching \cite{chen2015query,hazen2009query} that uses the dynamic time warping (DTW) algorithm to match the template by calculating distance. This method can save the training time but has poor robustness; (b) decoding method based on graph search \cite{wu2018monophone}  finds the best path on the decoding graph using the Viterbi algorithm, which still has strong competitiveness so far, but the cost of computation is expensive; (c) post-processing method \cite{chen2014small} computes the confidence score using a sliding window on the posterior probability. When the confidence score surpasses the threshold, the keyword is detected. This method has limited operations and is suitable for resource-constrained platforms. But it is worth noting that a neural network still needs to be pre-trained for frame-level alignment; (d) end-to-end method including sequence-to-sequence or end-to-end models \cite{zhuang2016unrestricted,alvarez2019end,he2017streaming}. The paper \cite{zhuang2016unrestricted} trained an LSTM model with connectionist temporal classification (CTC) for KWS that generates lattice for search. The paper \cite{sainath2015convolutional} trained a CNN for an end-to-end system to predict whether a keyword is spotted in an audio stream without decoding the audio into phonemes or word strings. 

\begin{figure}[th]
	\centering
	\includegraphics[width=0.9\columnwidth]{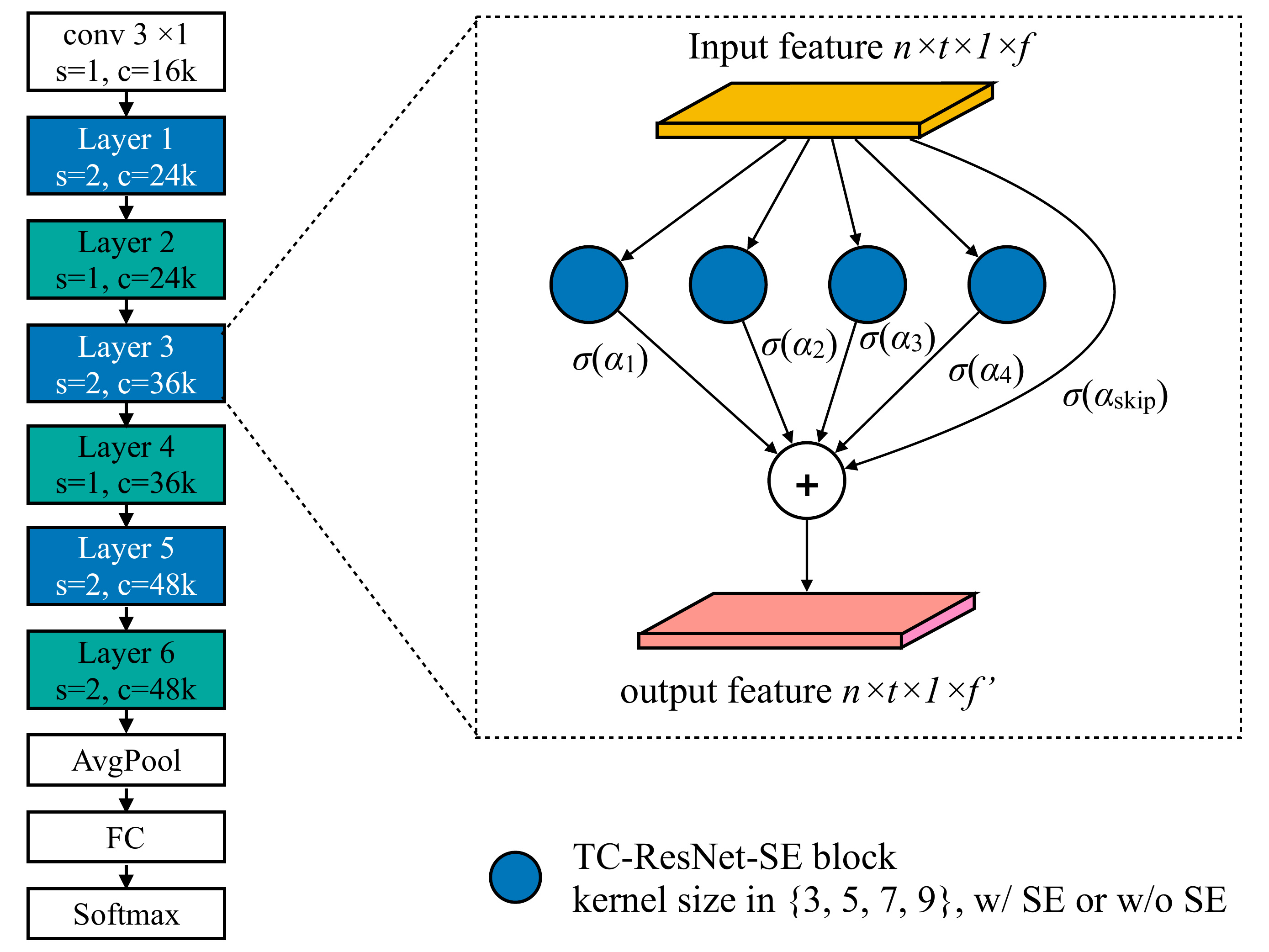}
	\caption{Differentiable architecture search in the adapted TC-ResNet search space. A searchable layer consists of several TC-ResNet-SE blocks and a skip connection, each is associated with an architectural parameter $\alpha$ to denote its importance. The outputs are summed up after multiplying $\sigma(\alpha)$. Note $\sigma$ can either be softmax for DARTS \cite{liu2018darts} and NoisyDARTS \cite{chu2020noisy}, or sigmoid for FairDARTS \cite{chu2019fair}. }
	\label{fig:tc-resnet-darts-supernet}
\end{figure}

An end-to-end KWS algorithm based on convolutional neural networks perceives the information of the entire window, which needs less assumptions and has large room for improvement. Therefore, this paper is based on this method to explore model architectures with higher performance.

\section{Prior Work}
\subsection{Manual Designed Networks for End-to-end KWS}
End-to-end KWS systems typically treat audio signals as images, either in the form of spectrograms or MFCCs. Keyword spotting is then converted into an image classification task. The paper \cite{zhang2017hello} explores various neural network architectures for KWS on embedded hardware in terms of accuracy, the number of parameters and operations, including the traditional DNN, CNN, LSTM, CRNN etc. The experimental results show that the traditional DNN has limited parameters and operations, but the accuracy is slightly worse; CNN has high accuracy but has a large number of operations; LSTM and CRNN can balance the number of operations and parameters and achieve good accuracy; MobileNetV1 achieves the best result  because its deep separable structure deepens the network and performs better. The paper \cite{tang2018deep} explores the application of deep residual learning and dilated convolutions to KWS and the model Res15 achieves 95.8\% accuracy and surpasses CNN \cite{sainath2015convolutional}. TC-ResNet \cite{choi2019temporal} proposed a \emph{temporal convolutional neural network} for real-time KWS on embedded hardware using 1D convolution along temporal dimension. TC-ResNet8 proposed in this paper achieves 385x speedup and the accuracy is improved 0.3\% accuracy compared to the deep and complex Res15 \cite{tang2018deep}. To reduce the number of parameters, \cite{mittermaier2020small} proposed a new architecture grouping depthwise separable convolutions (GDSConv) that achieves 96.4\% accuracy with 62k parameters. 

\subsection{Neural Architecture Search and Audio}
Neural architecture search (NAS) has already become a new paradigm of designing neural networks for many deep learning tasks. DARTS \cite{liu2018darts} tremendously reduces the search cost with a weight-sharing mechanism, which adopts gradient descent for a bi-level optimization on network parameters and architectural coefficients. It is however known to be unstable to reproduce. Our recent works FairDARTS \cite{chu2019fair} and NoisyDARTS \cite{chu2020noisy} study its failure case in-depth and propose multiple ways  to robustify DARTS. Specifically,  FairDARTS \cite{chu2019fair} breaks the exclusive competition among paralleling operations and imposes an auxiliary loss to push architectural coefficients towards its extremities.  NoisyDARTS \cite{chu2020noisy} attenuates the unfair advantage of skip connections by adding a small amount of Gaussian noise to their feature maps during the optimization. Both methods have been proved effective in classification tasks.

For audio tasks, NAS has also attracted considerable attention. Apart from our previous work NASC \cite{li2019neural} adopting our two-stage one-shot NAS approach FairNAS \cite{chu2019fairnas} on acoustic scene classification, DARTS \cite{chu2019fair} has been also applied to speaker recognition in AutoSpeech \cite{ding2020autospeech}, and to speech recognition in DARTS-ASR \cite{chen2020darts}. There is a noticeable contemporary work \cite{mo2020neural} also applying DARTS on KWS. However, due to the complex cell-based network topology, their searched networks might be limited for direct application on smart devices. Noticeably, there are some other NAS approaches on keyword spotting, such as NASIL \cite{fard2020nasil} and \cite{mazzawi2019improving}, however they are relatively costly (60 and 50 GPU hours respectively) and both generate less competitive results. 

\section{Method}

In this section, we undertake an efficient differentiable neural architecture search approach for the keyword spotting task. We first design a viable search space and then we perform DARTS and our robustified variants for searching.

\subsection{Search Space Design and Analysis}\label{sec:ss}

\begin{figure}[th]
	\centering
	\includegraphics[width=0.9\columnwidth]{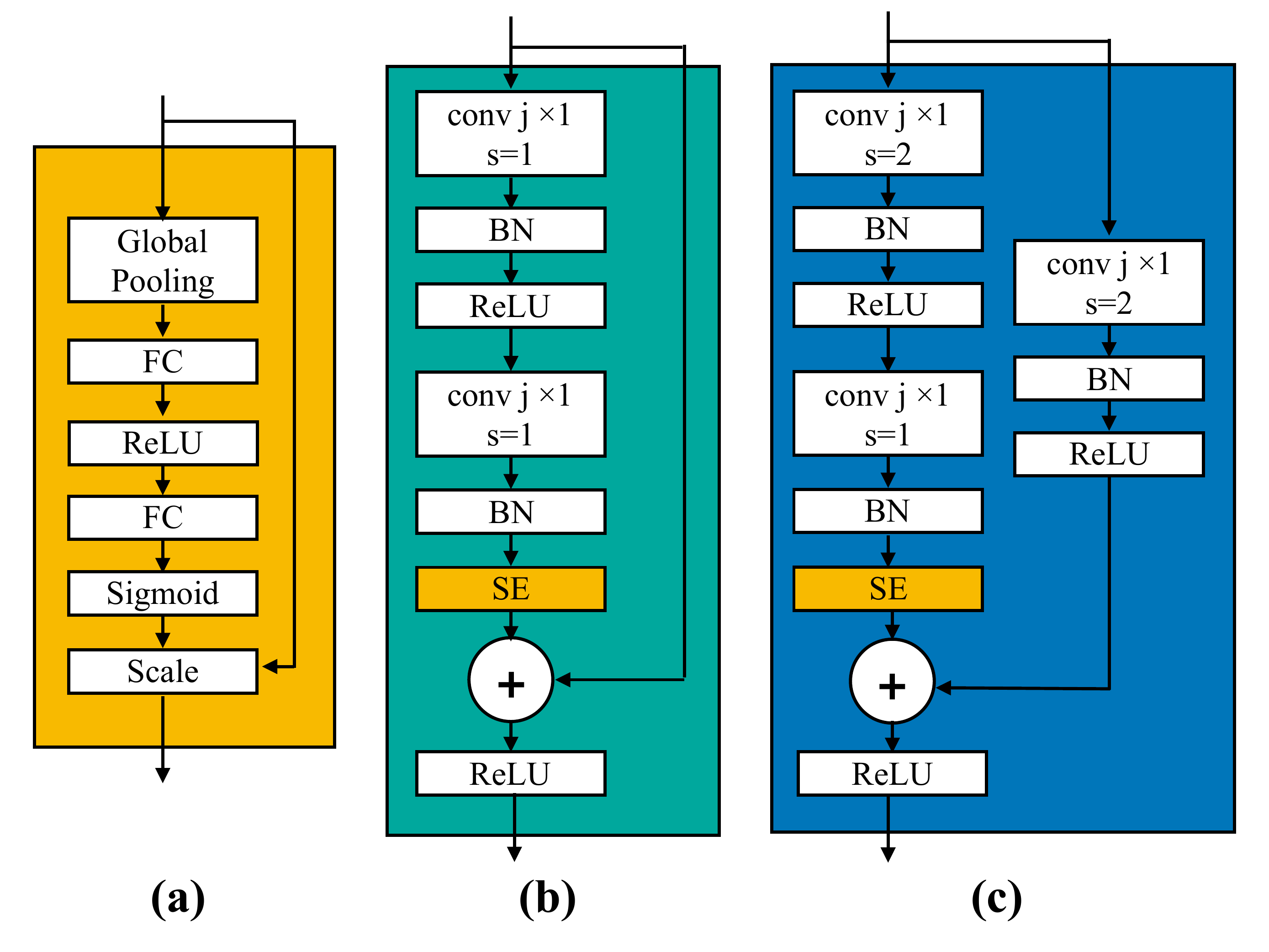}
	\caption{The searchable TC-ResNet-SE block (abbreviated as TC hereafter). The kernel size $j \in \{3,5,7,9\}$. (a) optional SE module, (b) normal block, (c) reduction block (stride s=2) }
	\label{fig:tc-resnet-se-block}
\end{figure}

We design our search space on top of TC-ResNet \cite{choi2019temporal} due to its outstanding performance and small memory footprint. We also introduce the squeeze-and-excitation (SE) module \cite{hu2018squeeze} for the TC-ResNet block, see Fig.~\ref{fig:tc-resnet-se-block}. To measure whether the search space is well set, we  manually fix convolutional kernels as \{3,5,7,9\} to train each model (trained 7 times and evaluated on the test set). The results are shown in Table~\ref{table:sota-google-scd}. We discover that under manual settings, the performance ranges from 95.97 \% to 96.98\% for TC-ResNet14-SE on V1 dataset and from 96.33\% to 97.22\% on V2. 

\begin{table}[h!]
\setlength{\tabcolsep}{3pt}
\centering
\small
\begin{tabular}{|c | c | c || c | c | } 
\hline
\multirow{2}{*}{Kernels} & \multicolumn{2}{c||}{V1} &  \multicolumn{2}{c|}{V2} \\
\cline{2-5}
  & TC14 & TC14-SE  & TC14 & TC14-SE \\ 
\hline
 3 &  96.64$\pm$0.20 &  96.64$\pm$0.18 &  96.89$\pm$0.22 &  96.94$\pm$0.19  \\ 
 5 &  96.66$\pm$0.21 &  96.60$\pm$0.20 &  96.78$\pm$0.13  &  96.96$\pm$0.16 \\
 7 & 96.45$\pm$0.25  &  96.55$\pm$0.11 &  96.66$\pm$0.18 &  96.64$\pm$0.29 \\
 9 & 96.49$\pm$0.18  &  96.52$\pm$0.25 &  96.79$\pm$0.18 & 96.78$\pm$0.19  \\ 
 \hline
\end{tabular}
\caption{Test accuracy of models in TC-ResNet search space with fixed kernels on Google Speech Command Dataset V1 and V2. SE means squeeze-and-excitation for each TC block. TC14 is short for TC-ResNet-14, the channel multiplier is set to 1.5. Each setting is run for 7 times to get the average and standard variation.}
\label{table:sota-google-scd}
\end{table}

Specifically, for each TC block we have options of kernel sizes in \{3,5,7,9\}, whether to enable SE or not, and an additional skip connection. In total, we have $6^9 \approx 10M $ models in the whole search space. 

\subsection{Searching Algorithm}
Considering the efficiency and effectiveness, we adopt DARTS \cite{liu2018darts}, FairDARTS \cite{chu2019fair} and NoisyDARTS \cite{chu2020noisy} for searching in the search space defined in Section~\ref{sec:ss}. Assume $\mathcal{O}$ be the set of candidate operations per layer. DARTS \cite{liu2018darts} assigns each candidate operation $o \in \mathcal{O}$ an architectural weight $\alpha_o$ and relaxes the discrete choices to be continuous. For our proposed search space, it constructs an over-parameterized network (see Fig.\ref{fig:tc-resnet-darts-supernet}) where the $i$-th layer's output is a weighted summation of the outputs of all candidate operations, i.e.,
\begin{align}\label{eq:darts-relax}
x_j = \sum_{o \in \mathcal{O}} \frac{e^{\alpha_o}}{\sum_{o' \in \mathcal{O}} e^{\alpha_{o'}} } o (x_i)
\end{align} 
To decide which operation should be chosen per layer, it formulates the searching process as a bi-level optimization problem,
\begin{align}\label{eq:darts-def}
 &min_{\alpha} \quad \mathcal{L}_{val}(w^*(\alpha), \alpha) \\
s.t. \quad &w^*(\alpha) = \arg \min_{w} \mathcal{L}_{train} (w, \alpha)
\end{align} 
where $\mathcal{L}_{val}$ and $\mathcal{L}_{train}$ denote the training and validation loss respectively. This is approximately solved by stochastic gradient descent (SGD) algorithm iteratively step by step. Specifically, the network weights $w$ and architectural weights $\alpha$ are updated in an interleaved order by two optimizers w.r.t. $\mathcal{L}_{val}$ and $\mathcal{L}_{train}$, each computed on the current batch per step.

We have shown how to construct different operations in the same layer in Figure~\ref{fig:tc-resnet-darts-supernet}. DARTS and FairDARTS differ in the activation function $\sigma$ used for architectural parameters $\alpha$. The former uses softmax (Equation \ref{eq:softmax}), and the latter goes with sigmoid (Equation \ref{eq:sigmoid}), which essentially makes each operation independent of others. Notice FairDARTS selects operations that are above a threshold ($\sigma=0.8$ in our case) while DARTS chooses the one with the highest $\sigma$.


\begin{equation}\label{eq:softmax}
softmax(\alpha_o) = \frac{e^{\alpha_o}}{\sum_{o' \in \mathcal{O}} e^{\alpha_{o'}} } \\
\end{equation}
\begin{equation}\label{eq:sigmoid}
sigmoid(\alpha_o) = \frac{1}{1+e^{-\alpha_o}}
\end{equation}

NoisyDARTS \cite{chu2020noisy} observes that skip connection causes performance collapse by forming up a residual structure \cite{he2016deep}. It thus injects Gaussian noise to perturb the fluent gradient flow. In short, it blends the skip connection's output with noise as follows,  where the mean $\mu$ is normally set to zero (to be unbiased) and the standard deviation $\beta$ to a small value.
\begin{equation}
f_{skip} (x) =  x + \mathcal{N}(\mu, \beta)
\end{equation} 

\section{Experiments}
\subsection{Google Speech Commands Dataset}
This paper explores the end-to-end neural network architecture on the Google speech commands datasets V1 and V2 \cite{warden2018speech}. With the standard splitting setup where the dataset was divided into a training set, a validation set and a test set with a ratio of 8:1:1 so that V1 contains 22246, 3093, 3081 samples for each set and V2 has 36923, 4445, 4890. Each audio is 1 second long and contains only one keyword. There are a total of 12 classes, including 10 keywords (Yes, No, Up, Down, Left, Right, On, Off, Stop, Go) and two extra classes: silence and unknown (sampled from the remaining 20 keywords). For the reliability of the experimental results, the audios of one person can only be divided into the same data set. 

\subsection{Searching}
For searching settings, we follow DARTS \cite{liu2018darts} with minor modifications. We train the supernet with a batch size of 128 for 50 epochs. We set a learning rate of 0.1 for the momentum SGD optimizer of network weights and 3e-4 for the Adam optimizer \cite{kingma2014adam} of architectural weights. We use additive Gaussian noise with $\beta$=0.1 for NoisyDARTS experiments on V1 and $\beta$=0.3 for V2. We use $w_{01}$=0.2 to balance entropy loss and $L_{01}$ in FairDARTS \cite{chu2019fair}. The searching takes about 2 hours on a single V100 GPU.

\subsection{Single Network Training}
The input of the neural network is a 40-dimensional MFCC with better effect and closer to the characteristics of the human ear. We follow the exact same setting as TC-ResNet \cite{choi2019temporal} to generate MFCCs and to train all the models. It takes nearly a hour on a single V100 GPU. The evaluation metric is top-1 accuracy calculated on the splitted test dataset.


\subsection{Searching Results}

\begin{figure}[ht]
	\centering
	\includegraphics[width=0.98\columnwidth]{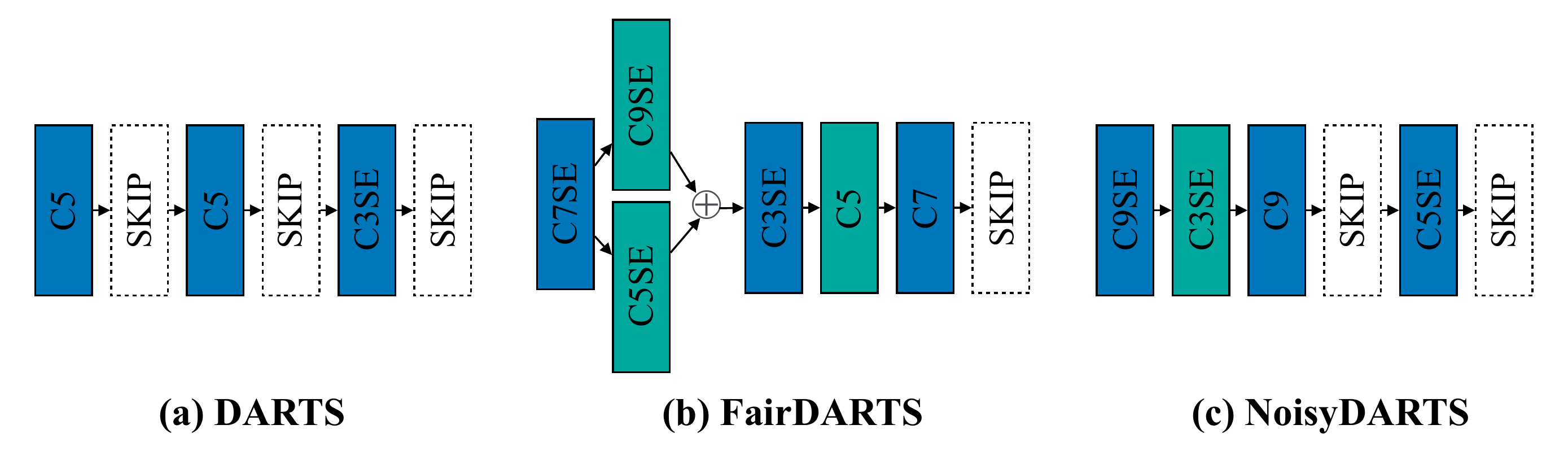}
	\caption{Best models found by DARTS and its variants on v1 dataset, the stem and the tail are omitted. Note C$x$SE means a TC-ResNet block with kernel size $x$ and it has an SE module. Dark blue indicates downsampling with stride $s=2$. For dark green layers,  $s=1$.}
	\label{fig:v1-best-models}
\end{figure}

After searching with the adopted methods, we fully train each model from scratch. The results are shown in Table~\ref{table:sota-google-scd-v1} and Table~\ref{table:sota-google-scd-v2}. The best models are visualized in Figure~\ref{fig:v1-best-models}. The searching and evaluation are proxyless either on V1 or V2 dataset. We refrain from using techniques like SpecAugment \cite{park2019specaugment} and self-attention \cite{rybakov2020streaming} to have a fair comparison with other prior arts. Notice that with these tricks, the performance can be further boosted.

From the experiment results, we find that DARTS tends to select many skip connections (e.g. Fig.~\ref{fig:v1-best-models}) so that the derived model has much less parameters, also less competitive in terms of accuracy. This is a known issue in cell-based search spaces as well, which is considered to be a result of unfair competition by forming a residual structure \cite{chu2019fair}. FairDARTS alleviates this problem by allowing multiple choices per layer, see (b) in Figure~\ref{fig:v1-best-models}. NoisyDARTS perturbs the output from skip connections with noise injection to block the gradient flow from the residual structure. By controlling the standard deviation $\beta$ of the noise, we can tune the searching algorithm to find a trade-off between the number of skip connections and its overall performance. NoisyDARTS finds the best model out of all three methods on V1 dataset, whose average number of parameters is nearly \textbf{8$\times$} fewer than the contemporary work NAS2 \cite{mo2020neural}. With much improved efficiency, it allows us to deploy our models on IoT devices with low computation consumption. As random search is considered to be a powerful baseline again NAS methods, we also random sample architectures from the design search space which renders models (Random-TC14) that are slightly worse than DARTS, but not good enough to outperform NoisyDARTS. It is although interesting to see all three DARTS methods agree to have a skip connection in the last layer.

\begin{table}[ht]
\centering
\begin{small}
\setlength\tabcolsep{4pt} 
\begin{tabular}{|c | c | c | c | c |} 
 \hline
 Method & Params & $\times+$ & Avg. acc (\%) & Best  \\ [0.5ex] 
 \hline
 TC-ResNet-14 \cite{choi2019temporal} &  305K & 13.4M &   96.49$\pm$0.18$^\star$ & 96.7$^\star$  \\ 
 CENet-GCN-40 \cite{chen2019small} &  72.3K  & 16.18M & 96.8 & 97.0 \\
MHAtt-RNN \cite{rybakov2020streaming} & 743K$^\star$ & 87.2M$^\star$ & - & 97.2$^\dagger$ \\
NAS2 \cite{mo2020neural} & 886K & - & - &  97.2 \\
Random-TC14$^\ddagger$ & 196K & 8.8M & 96.58$\pm$0.15 & 96.8 \\
DARTS-TC14$^\ddagger$ &  93K & 4.9M & 96.63$\pm$0.22 & 96.9  \\
FairDARTS-TC14$^\ddagger$ & 188K  & 10.6M  & 96.70$\pm$0.11 & 96.9  \\
NoisyDARTS-TC14$^\ddagger$ & 109K  & 6.3M & 96.79$\pm$0.30 & 97.2  \\ [1ex] 
 \hline
\end{tabular}
\end{small}
\caption{Comparison with state-of-the-art lightweight models on Google Speech Command Dataset v1. $^\star$: rerun/tested from their source code. $^\dagger$: w/ SpecAugment. $^\ddagger$: Averaged on 8 searched models.}
\label{table:sota-google-scd-v1}
\end{table}

Results on V2 dataset in Table \ref{table:sota-google-scd-v2} again attests that our approach can stably find competitive networks compared with human-designed baselines. Performance on V2 is generally better than that of V1 since it has more data samples which benefits both the searching and the evaluation process. It also confirms that DARTS variants have improved searching capability since the defects of DARTS are effectively solved by FairDARTS and NoisyDARTS.

\begin{table}[ht]
\centering
\begin{small}
\setlength\tabcolsep{4pt} 
\begin{tabular}{|c | c | c | c | c |} 
 \hline
 Method & Params & $\times+$ & Avg. acc (\%) & Best  \\ [0.5ex] 
 \hline
 TC-ResNet-14 \cite{choi2019temporal} &  305K & 13.4M &   96.79$\pm$0.18$^\star$ & 97.03$^\star$  \\ 
DARTS-TC14$^\ddagger$ & 88K  &  5.5M & 96.92$\pm$0.23   & 97.14   \\
FairDARTS-TC14$^\ddagger$ & 156K   & 8.5M & 97.11$\pm$0.14  &  97.35 \\
NoisyDARTS-TC14$^\ddagger$ & 107K  & 6.3M & 97.18$\pm$0.26 & 97.44 \\ [1ex]  
 \hline
\end{tabular}
\end{small}
\caption{Comparison with state-of-the-art lightweight models on Google Speech Command Dataset v2. $^\star$: train a single model or 8 times. $^\ddagger$: Averaged on 8 searched models.}
\label{table:sota-google-scd-v2}
\end{table}

\begin{figure}[ht]
	\centering
	\includegraphics[width=0.7\columnwidth]{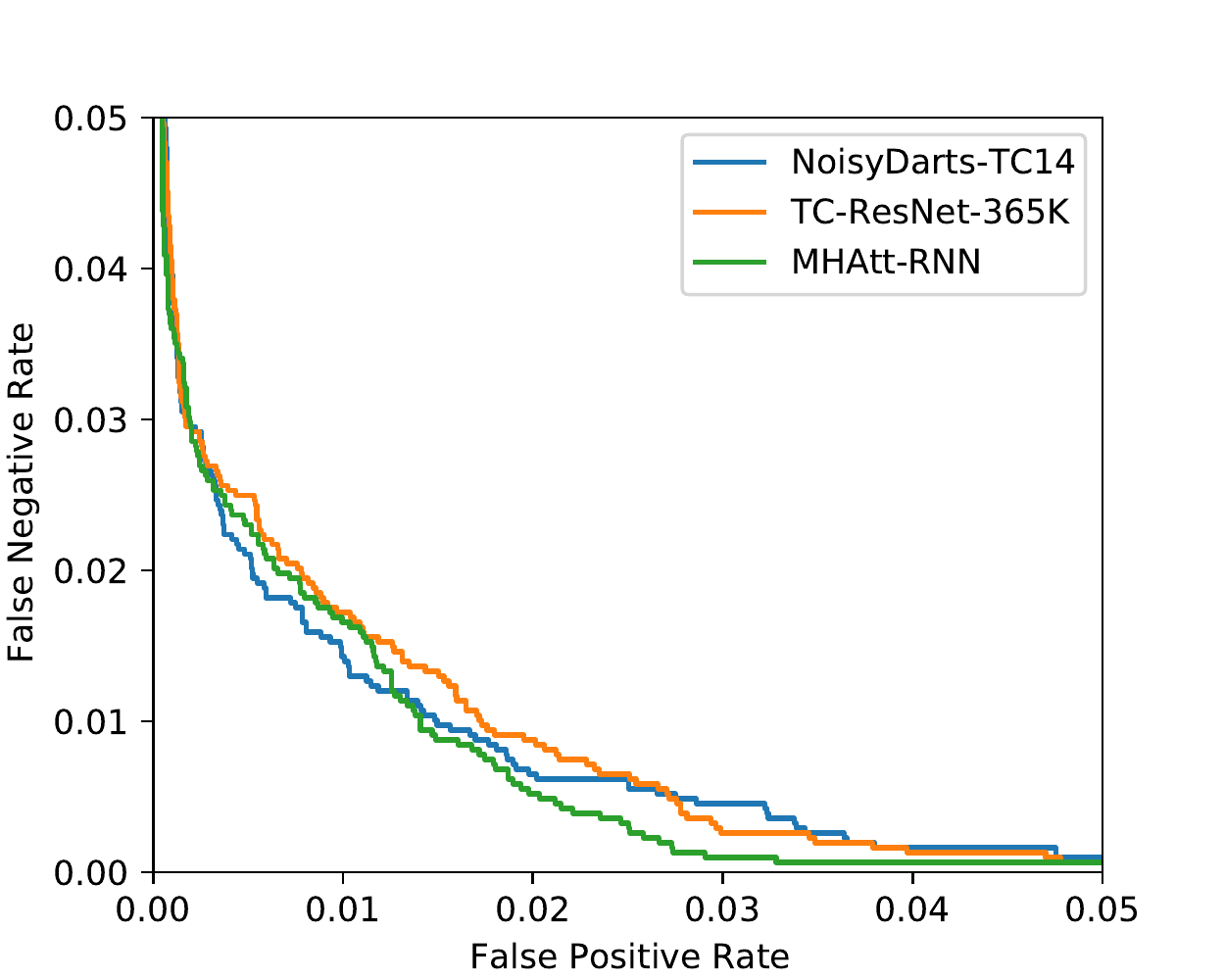}
	\caption{ROC curve of selected models of the similar accuracy on V1 dataset.}
	\label{fig:roc-curve}
\end{figure}

We take NoisyDARTS model trained on V1 dataset and plot the ROC curve with false negative rate vs. false positive rate in Figure~\ref{fig:roc-curve}, in comparison with MHAtt-RNN and TC-ResNet-365K (boosted version with 365K parameters by \cite{rybakov2020streaming}). It indicates that NoisyDARTS-TC-14 and MHAtt-RNN are close, while both outperforming TC-ResNet-365K.

\subsection{Discussion}
Albeit the advancement by the proposed method, there is still space for improvement, i.e., to test other DARTS variants, to take latency or multiply-adds into account during searching, to design a richer search space such as variable channel numbers and inter-block connections etc. 



\section{Conclusion}
This paper exploits the differentiable neural architecture search on the keyword spotting task with practical reasoning in mind. Due to its strict hardware constraints and high performance requirement, we designed an efficient and applicable search space from TC-ResNet with minor modifications. We investigated DARTS and its two variants FairDARTS and NoisyDARTS for searching. The found architectures achieve state-of-the-art results on the Google Speech Command Dataset, with nearly 8$\times$ fewer number of parameters compared with prior art. We hope this effort could give a new paradigm to the future architecture design for audio systems.


\let\OLDthebibliography\thebibliography
\renewcommand\thebibliography[1]{
  \OLDthebibliography{#1}
  \setlength{\parskip}{0pt}
  \setlength{\itemsep}{0pt plus 0.3ex}
}

\bibliographystyle{IEEEbib}
\bibliography{refs}

\begin{thebibliography}{10}

\bibitem{chen2015query}
Guoguo Chen, Carolina Parada, and Tara~N Sainath,
\newblock ``Query-by-example keyword spotting using long short-term memory
  networks,''
\newblock in {\em ICASSP}. IEEE, 2015, pp. 5236--5240.

\bibitem{hazen2009query}
Timothy~J Hazen, Wade Shen, and Christopher White,
\newblock ``Query-by-example spoken term detection using phonetic posteriorgram
  templates,''
\newblock in {\em ASRU Workshop}. IEEE, 2009, pp. 421--426.

\bibitem{wu2018monophone}
Minhua Wu, Sankaran Panchapagesan, Ming Sun, Jiacheng Gu, Ryan Thomas, Shiv
  Naga~Prasad Vitaladevuni, Bjorn Hoffmeister, and Arindam Mandal,
\newblock ``Monophone-based background modeling for two-stage on-device wake
  word detection,''
\newblock in {\em ICASSP}. IEEE, 2018, pp. 5494--5498.

\bibitem{chen2014small}
Guoguo Chen, Carolina Parada, and Georg Heigold,
\newblock ``Small-footprint keyword spotting using deep neural networks,''
\newblock in {\em ICASSP}. IEEE, 2014, pp. 4087--4091.

\bibitem{zhuang2016unrestricted}
Yimeng Zhuang, Xuankai Chang, Yanmin Qian, and Kai Yu,
\newblock ``Unrestricted vocabulary keyword spotting using lstm-ctc,''
\newblock in {\em Interspeech}, 2016, pp. 938--942.

\bibitem{alvarez2019end}
Raziel Alvarez and Hyun-Jin Park,
\newblock ``End-to-end streaming keyword spotting,''
\newblock in {\em ICASSP}. IEEE, 2019, pp. 6336--6340.

\bibitem{he2017streaming}
Yanzhang He, Rohit Prabhavalkar, Kanishka Rao, Wei Li, Anton Bakhtin, and Ian
  McGraw,
\newblock ``Streaming small-footprint keyword spotting using
  sequence-to-sequence models,''
\newblock in {\em ASRU Workshop}. IEEE, 2017, pp. 474--481.

\bibitem{sainath2015convolutional}
Tara~N Sainath and Carolina Parada,
\newblock ``Convolutional neural networks for small-footprint keyword
  spotting,''
\newblock in {\em ISCA}, 2015.

\bibitem{liu2018darts}
Hanxiao Liu, Karen Simonyan, and Yiming Yang,
\newblock ``{DARTS: Differentiable architecture search},''
\newblock in {\em ICLR}, 2019.

\bibitem{chu2020noisy}
Xiangxiang Chu, Bo~Zhang, and Xudong Li,
\newblock ``Noisy differentiable architecture search,''
\newblock {\em arXiv preprint arXiv:2005.03566}, 2020.

\bibitem{chu2019fair}
Xiangxiang Chu, Tianbao Zhou, Bo~Zhang, and Jixiang Li,
\newblock ``{Fair DARTS: Eliminating unfair advantages in differentiable
  architecture search},''
\newblock in {\em ECCV}, 2020.

\bibitem{zhang2017hello}
Yundong Zhang, Naveen Suda, Liangzhen Lai, and Vikas Chandra,
\newblock ``Hello edge: Keyword spotting on microcontrollers,''
\newblock {\em arXiv preprint arXiv:1711.07128}, 2017.

\bibitem{tang2018deep}
Raphael Tang and Jimmy Lin,
\newblock ``Deep residual learning for small-footprint keyword spotting,''
\newblock in {\em ICASSP}. IEEE, 2018, pp. 5484--5488.

\bibitem{choi2019temporal}
Seungwoo Choi, Seokjun Seo, Beomjun Shin, Hyeongmin Byun, Martin Kersner,
  Beomsu Kim, Dongyoung Kim, and Sungjoo Ha,
\newblock ``Temporal convolution for real-time keyword spotting on mobile
  devices,''
\newblock {\em arXiv preprint arXiv:1904.03814}, 2019.

\bibitem{mittermaier2020small}
Simon Mittermaier, Ludwig K{\"u}rzinger, Bernd Waschneck, and Gerhard Rigoll,
\newblock ``Small-footprint keyword spotting on raw audio data with
  sinc-convolutions,''
\newblock in {\em ICASSP}. IEEE, 2020, pp. 7454--7458.

\bibitem{li2019neural}
Jixiang Li, Chuming Liang, Bo~Zhang, Zhao Wang, Fei Xiang, and Xiangxiang Chu,
\newblock ``Neural architecture search on acoustic scene classification,''
\newblock in {\em INTERSPEECH}, 2020.

\bibitem{chu2019fairnas}
Xiangxiang Chu, Bo~Zhang, Ruijun Xu, and Jixiang Li,
\newblock ``Fairnas: Rethinking evaluation fairness of weight sharing neural
  architecture search,''
\newblock {\em arXiv preprint arXiv:1907.01845}, 2019.

\bibitem{ding2020autospeech}
Shaojin Ding, Tianlong Chen, Xinyu Gong, Weiwei Zha, and Zhangyang Wang,
\newblock ``Autospeech: Neural architecture search for speaker recognition,''
\newblock {\em arXiv preprint arXiv:2005.03215}, 2020.

\bibitem{chen2020darts}
Yi-Chen Chen, Jui-Yang Hsu, Cheng-Kuang Lee, and Hung-yi Lee,
\newblock ``{DARTS-ASR: Differentiable Architecture Search for Multilingual
  Speech Recognition and Adaptation},''
\newblock {\em arXiv preprint arXiv:2005.07029}, 2020.

\bibitem{mo2020neural}
Tong Mo, Yakun Yu, Mohammad Salameh, Di~Niu, and Shangling Jui,
\newblock ``{Neural Architecture Search For Keyword Spotting},''
\newblock in {\em Interspeech}, 2020.

\bibitem{fard2020nasil}
Farzaneh~S Fard, Arash Rad, and Vikrant~Singh Tomar,
\newblock ``Nasil: Neural architecture search with imitation learning,''
\newblock in {\em ICASSP}. IEEE, 2020, pp. 3972--3976.

\bibitem{mazzawi2019improving}
Hanna Mazzawi, Xavi Gonzalvo, Aleks Kracun, Prashant Sridhar, Niranjan
  Subrahmanya, Ignacio Lopez-Moreno, Hyun-Jin Park, and Patrick Violette,
\newblock ``{Improving Keyword Spotting and Language Identification via Neural
  Architecture Search at Scale},''
\newblock in {\em INTERSPEECH}, 2019, pp. 1278--1282.

\bibitem{hu2018squeeze}
Jie Hu, Li~Shen, and Gang Sun,
\newblock ``{Squeeze-and-excitation networks},''
\newblock in {\em CVPR}, 2018, pp. 7132--7141.

\bibitem{he2016deep}
Kaiming He, Xiangyu Zhang, Shaoqing Ren, and Jian Sun,
\newblock ``Deep residual learning for image recognition,''
\newblock in {\em CVPR}, 2016, pp. 770--778.

\bibitem{warden2018speech}
Pete Warden,
\newblock ``Speech commands: A dataset for limited-vocabulary speech
  recognition,''
\newblock {\em arXiv preprint arXiv:1804.03209}, 2018.

\bibitem{kingma2014adam}
Diederik~P Kingma and Jimmy Ba,
\newblock ``Adam: A method for stochastic optimization,''
\newblock {\em arXiv preprint arXiv:1412.6980}, 2014.

\bibitem{park2019specaugment}
Daniel~S Park, William Chan, Yu~Zhang, Chung-Cheng Chiu, Barret Zoph, Ekin~D
  Cubuk, and Quoc~V Le,
\newblock ``{SpecAugment: A simple data augmentation method for automatic
  speech recognition},''
\newblock in {\em Interspeech}, 2019.

\bibitem{rybakov2020streaming}
Oleg Rybakov, Natasha Kononenko, Niranjan Subrahmanya, Mirko Visontai, and
  Stella Laurenzo,
\newblock ``{Streaming keyword spotting on mobile devices},''
\newblock {\em arXiv preprint arXiv:2005.06720}, 2020.

\bibitem{chen2019small}
Xi~Chen, Shouyi Yin, Dandan Song, Peng Ouyang, Leibo Liu, and Shaojun Wei,
\newblock ``Small-footprint keyword spotting with graph convolutional
  network,''
\newblock in {\em ASRU Workshop}. IEEE, 2019, pp. 539--546.

\end{thebibliography}

\end{document}